\documentclass[aps,prl,twocolumn]{revtex4-1}
\usepackage{graphics,graphicx}
\usepackage{color}
\usepackage{wrapfig}
\usepackage{enumerate}
\usepackage{amssymb,amsmath}
\usepackage{bm}
\usepackage[normalem]{ulem} 
\newcommand{\red}[1]{}\newcommand{\blue}{} 

\begin{document}
\title{Quenching the Haldane gap in spin-1 Heisenberg antiferromagnets}
\author{Keola~Wierschem$^{1,2}$ and Pinaki~Sengupta$^{1}$}
\affiliation{$^{1}$School of Physical and Mathematical Sciences, Nanyang Technological University, 21 Nanyang Link, Singapore 637371\\$^{2}$International Institute for Complex Adaptive Matter, University of California, Davis, CA 95616}
\date{\today}
\begin{abstract}
We consider a quasi-one-dimensional system of spin-1 Heisenberg antiferromagnetic chains in 2D and 3D hypercubic lattices with interchain coupling $J$ and uniaxial single-ion anisotropy $D$. Using large scale numerical simulations, we map out the $J-D$ phase diagram and investigate the low lying excitations of the Haldane phase in the $J\ll1$ limit. We also provide direct evidence that the Haldane phase remains a non-trivial symmetry protected topological state for small but finite $J$.
\end{abstract}
\maketitle


{\em Introduction.---}Enhanced quantum fluctuations drive several novel quantum phases  in interacting quantum spins in low dimensions -- phases that are suppressed in higher dimensions. The search for such phases and the quest for understanding the mechanism behind their emergence has kept the study of quantum magnetism one of the most active frontiers of Condensed Matter Physics. The confluence of continuing experimental advances in preparing and characterizing low dimensional quantum magnets and the simultaneous development of powerful analytic and numeric methods analyzing relevant microscopic models has resulted in great advancement of our understanding of the many-body effects underlying these unique states of matter. A striking result in the study of quantum spin systems is the pioneering work by Haldane~\cite{Haldane1983a,*Haldane1983b}. By studying the non-linear sigma model in (1+1) dimensions, Haldane conjectured that the ground state of the one-dimensional (1D) Heisenberg antiferromagnet (HAFM) has gapless excitations for half-odd integer spins, whereas that for integer spins is separated from all excited states by a finite spin gap (Haldane gap). Haldane's conjecture has inspired numerous theoretical studies of integer spins in low dimensions, including chain mean field theory (CMFT)~\cite{Sakai1989,Sakai1990}, exact diagonalization~\cite{Botet1983,*Golinelli1992}, density matrix renormalization group (DMRG)~\cite{White1993,*White2008,Hu2011,Moukouri2011,*Moukouri2012}, and quantum Monte Carlo~(QMC) simulations~\cite{Albuquerque2009,Kim2000,Matsumoto2001,Wierschem2012,Wierschem2013}. Most of these studies have focused on $S=1$ spins where the Haldane gap is the most robust.

The theoretical studies have been complemented by the discovery of several quasi-one-dimensional (Q1D) spin $S=1$ quantum magnets, such as AgVP$_2$S$_6$~\cite{Mutka1991,*Asano1994,*Takigawa1995,*Takigawa1996}, NDMAP~\cite{Honda2001}, NENP~\cite{Renard1987,*Renard1988,*Regnault1994,*Zaliznyak1998}, NINO~\cite{Renard1988}, PbNi$_2$V$_2$O$_8$~\cite{Uchiyama1999,*Zheludev2000}, SrNi$_2$V$_2$O$_8$~\cite{Pahari2006}, TMNIN~\cite{Gadet1991}, and Y$_2$BaNiO$_5$~\cite{Darriet1993,*Xu1996}. In these materials, the magnetic ions are arranged in chains, with weak but finite inter-chain couplings, which affect the ground state phases. Additionally, in most  known $S=1$ magnets, the ubiquitous Heisenberg exchange is complemented by a single-ion anisotropy. The expanded Hilbert space of the $S=1$ spins, and the interplay between multiple competing interactions, external magnetic field and different lattice geometries result in a rich variety of ground state phases. In addition to the gapped Haldane phase, examples of exotic quantum states realized in low dimensional interacting spin systems include experimentally realized Bose Einstein Condensation~(BEC) of magnons~\cite{Ruegg2004}, quantum paramagnet~\cite{Zapf2006,*Yin2008}, and the recently proposed spin supersolid~\cite{Sengupta2007a,*Sengupta2007b} and ferronematic~\cite{Wierschem2012f} phases. Recent advances in synthesis techniques has made it possible to engineer quasi-low-dimensional materials where the ``effective dimensionality'' (that is, inter-chain or inter-layer couplings) and Hamiltonian parameters (such as the ratio of exchange interaction and single-ion anisotropy) can be controlled. This raises the possibility of preparing materials with desired pre-determined properties. The search for such tailor-made materials has grown in recent years as these are believed to drive the next generation of electronics. In addition to condensed matter systems, rapid advances in the field of ultracold atoms in optical lattices have opened up a new frontier in the study of interacting many-body systems in arbitrary dimensions. The unprecedented control over number of atoms, interactions, and lattice geometry makes it an ideal testbed for preparing and studying novel quantum states.

In this work, we study the $S=1$ HAFM with single-ion anisotropy on Q1D lattices in 2D and 3D, focusing on the behavior of the Haldane phase as the different Hamiltonian parameters are varied. Previous QMC~\cite{Kim2000,Matsumoto2001,Wierschem2012,Wierschem2013} and DMRG~\cite{Moukouri2011,*Moukouri2012} studies of the isotropic $S=1$ HAFM in 2D and 3D have shown that the Haldane phase persists in the presence of small, but non-zero inter-chain couplings. The combined effect of inter-chain couplings and single-ion anisotropy has been studied using CMFT~\cite{Sakai1990}, but enhanced quantum fluctuations in the Q1D limit make the predictions from mean field theories unreliable. \red{\sout{However}}\blue{Further}, a detailed exploration of the ground state properties of the Haldane phase in this system, including the low lying excitation spectrum, does not exist to the best of our knowledge. Such a study is important to the understanding of experimental results for real quantum magnets. Additionally, the topological nature of the Q1D Haldane phase is still an open question. While the Haldane phase of the spin-1 chain is perhaps the simplest example of a state with symmetry protected topological (SPT) order~\cite{Gu2009,*Chen2013}, the ground state of the spin-1 ladder has been shown to be a topologically trivial state~\cite{Pollmann2012}. Here, we determine the ground state phase diagram of a system of weakly coupled \red{\sout{spin }}$S=1$ HAFM chains with uniaxial single-ion anisotropy. Further, we investigate the quantum phase transitions out of the Haldane phase and show the evolution of the low lying excitation spectrum with the closing of the Haldane gap. Finally, we present direct evidence of non-trivial SPT order in \red{\sout{this Q1D system}}\blue{the Q1D Haldane phase as defined in this system~\footnote{We define the Q1D Haldane phase as the gapped symmetric ground state of the model Hamiltonian~(\ref{hamiltonian}) with $J\neq0$ that is adiabatically connected to the Haldane phase in 1D as $J\rightarrow0$}}.


{\em Model and Methods.---}We study a spatially anisotropic $S=1$ HAFM consisting of chains in 2D and 3D hypercubic lattice geometries, as described by the Hamiltonian
\begin{equation}
{\cal H}=\sum_{\left<ij\right>_{\parallel}}\vec{S_{i}}\cdot\vec{S_{j}}
		+J\sum_{\left<ij\right>_{\perp}}\vec{S_{i}}\cdot\vec{S_{j}}
		+D\sum_i\left(S_i^z\right)^2.
\label{hamiltonian}
\end{equation}
Here we have set the spin coupling along the chains to unity, thereby defining the energy scale of our system. This leaves the interchain coupling $J$ and the single-ion anisotropy $D$ as our remaining Hamiltonian parameters. Simulation cells in 2D and 3D have dimensions $L_\perp\times L$ and $L_\perp\times L_\perp\times L$, respectively, with chain length $L$ and aspect ratio $R=L/L_\perp$ ranging from 4 to 8. Due to the spatial anisotropy of this model, cells with $R>1$ more rapidly approach the thermodynamic limit~\cite{Sandvik1999}. Unless otherwise stated, results are obtained on a 3D lattice.

To investigate the above model, we use the stochastic series expansion (SSE) QMC method~\cite{Sandvik1991} with directed loops~\cite{Syljuasen2002,*Syljuasen2003}. Within this formalism, the spin stiffness $\rho_s$ is easily obtained in terms of the global winding numbers~\cite{Sandvik1997}. At a critical point between gapped and gapless phases in $d$ dimensions, $\rho_s\sim L^{2-(d+z)}$ where $z$ is the dynamic critical exponent~\cite{Sandvik1998}. Thus, the crossing point of $\rho_sL^{(d+z)-2}$ for different system sizes is an estimate of the critical point. A similar scaling can be derived for the staggered magnetization at the boundary of an Ising antiferromagnet: $m_s^2\sim L^{2-(d+z)-\eta}$. These finite size scaling (FSS) forms can be used to accurately determine quantum phase boundaries. In such cases, we must keep a fixed aspect ratio, not only of the spatial dimensions, but also of the inverse temperature (i.e. $\beta\propto L^z$). This guarantees that we approach the ground state thermodynamic limit as $L\rightarrow\infty$.

In addition to the SSE method, we implement a projective QMC method that allows us to access both ground state expectation values as well as wave function overlaps between the ground state and a trivial product state. Details are given in the Supplementary Material~\cite{Wierschem2014}.

\begin{figure}
\centering
\includegraphics[clip,trim=0cm 0cm 0cm 3.85cm,width=\linewidth]{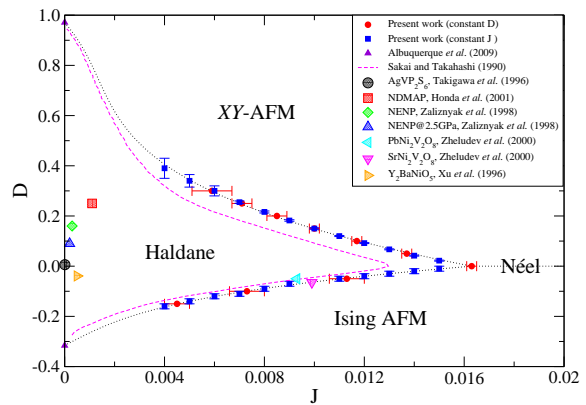}
\caption{(Color online) Phase diagram in the $J-D$ plane with phase boundaries as indicated by the dotted black lines. The borders of the Haldane phase are obtained as fits to the present work and represent guides for the eye. Data points are determined by QMC simulations at constant $D$ and $J$ (red circles and blue squares, respectively). For comparison, we show the CMFT results of Sakai and Takahashi~\cite{Sakai1990} as a dashed magenta line, while purple triangles represent QMC results of Albuquerque {\it et al.}~\cite{Albuquerque2009} for a 1D chain.}
\label{phases}
\end{figure}

{\em Phase Diagram.---}The ground state phase diagram of ${\cal H}$ is shown in Fig.~\ref{phases}. For small $|D|$ and $J$ the system is in the Haldane phase. For sufficiently strong interchain couplings, the Haldane gap is quenched and three dimensional long range magnetic order sets in. This magnetic order is the N\'{e}el antiferromagnetic state in the case of isotropic spins ($D=0$), while axial ($D<0$) and planar ($D>0$) anisotropy lead to Ising antiferromagnetic (Ising AFM) and $XY$ antiferromagnetic ($XY$-AFM) states, respectively. Additionally, there is a quantum paramagnetic phase for large $D\gtrsim1$ (not shown). The Haldane to $XY$-AFM phase boundary is determined by FSS of the spin stiffness, while the Haldane to Ising AFM phase boundary is determined by FSS of the staggered magnetization. In the case of the Haldane to N\'{e}el phase transition at the isotropic point ($D=0$), FSS of both the spin stiffness and staggered magnetization yield values in agreement up to the statistical uncertainty given.

The first thing to notice in the phase diagram of Fig.~\ref{phases} is the striking qualitative agreement between the CMFT results of Sakai and Takahashi~\cite{Sakai1990} and the QMC results of the current work. Although CMFT begins from exact results within each chain, it is expected to yield a lower bound for the extent of the Haldane phase. Our results confirm this expectation, and highlight that the difference between our unbiased QMC results and CMFT is largest at the isotropic point, where spin fluctuations are enhanced by the full SU(2) spin rotational symmetry.

To make contact with experiment, we plot the location of several Haldane gap materials in the phase diagram of Fig.~\ref{phases}. The tetragonal compounds PbNi$_2$V$_2$O$_8$ and SrNi$_2$V$_2$O$_8$ are of particular interest to our study. Due to their crystal symmetry, they are well described by a uniaxial crystal field. In addition, both lie near the boundary of the Haldane phase due to a combination of easy axis single ion anisotropy and interchain spin exchange coupling strength. In fact, SrNi$_2$V$_2$O$_8$ was originally believed to magnetically order below $T_N=7K$~\cite{Uchiyama1999,*Zheludev2000} based on experiments on powder samples, while more recent results on polycrystalline~\cite{Pahari2006} and single crystal\red{\sout{s}}~\cite{Bera2013} \blue{samples} are consistent with a non-magnetic \red{\sout{Haldane }}ground state. The proximity of SrNi$_2$V$_2$O$_8$ to a quantum phase boundary is made quite clear in the phase diagram of Fig.~\ref{phases}.

\begin{figure}
\centering
\includegraphics[clip,trim=0cm 0cm 0cm 2.75cm,width=\linewidth]{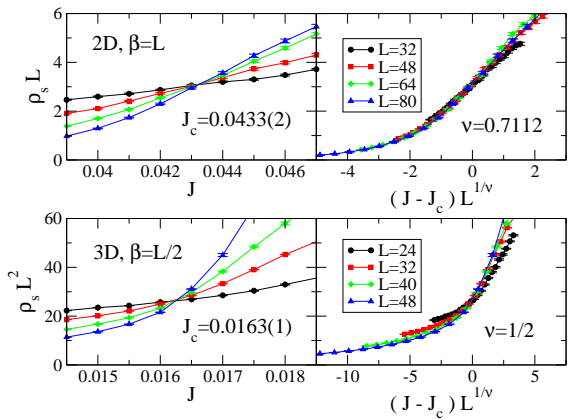}
\caption{(Color online) Finite size scaling of the spin stiffness $\rho_s$ in the Q1D limit of 2D and 3D lattices (upper and lower panels, respectively). Critical couplings \red{\sout{$J_c^{2{\text D}}=0.04333(2)$ and $J_c^{3{\text D}}=0.0163(1)$ }}are determined by the crossing criterion (left panels) assuming a dynamic critical exponent $z=1$. In both cases we use a single-ion anisotropy $D=0$ and aspect ratio $R=4$. The right panels demonstrate curve collapse around the critical points  using the critical exponents \red{\sout{$\nu\approx0.7112$~\cite{Campostrini2002} and $\nu=0.5$ of the Heisenberg universality class in 2+1 and 3+1 dimensions, respectively}}\blue{of the Heisenberg universality class~\cite{Campostrini2002}}.}
\label{scaling}
\end{figure}

{\em Spin Stiffness Scaling.---}To illustrate the spin stiffness scaling described above, we consider the ground state phase transition between the gapped Haldane phase and the gapless N\'{e}el state as the interchain coupling $J$ is varied at the isotropic point. For this quantum phase transition we expect \red{\sout{$d+z$ }}Heisenberg universality with $z=1$. In Fig.~\ref{scaling} we plot $\rho_sL^{d-1}$ for 2D and 3D lattices, along with the FSS collapse assuming critical exponents of the \red{\sout{$d+1$ }}Heisenberg universality class \blue{in $d+1$ dimensions}.

Our result for the critical coupling $J_c$ in 2D agrees well with past QMC~\cite{Kim2000,Matsumoto2001} and DMRG~\cite{Moukouri2011,*Moukouri2012} results. As expected, $J_c$ in 3D is much smaller, which is due to the larger role that fluctuations play in 2D as compared to 3D. This remains true even if we scale the results by the chain coordination number $n$ ($n=2$ for 2D and $n=4$ for 3D). In both cases, our scaled results are larger than the CMFT value $nJ_c\approx0.051$ that acts as a lower bound~\cite{Sakai1989}. Interestingly, QMC studies of different lattice geometries in 3D have found the mean field universality of $nJ_c$ to hold quite well for unfrustrated lattices~\cite{Wierschem2013}.


{\em Low Lying Excitations.---}We estimate the spin gap excitations $\omega_{\bf k}$ using the upper bound estimator $2S_{\bf k}/\chi_{\bf k}\ge\omega_{\bf k}$~\cite{Sandvik1996,*Wang2006}. In the Haldane phase this estimator is expected to perform well near $k_z=\pi$ due to the sharp nature of the single-magnon peak as well as an additional gap to the multi-magnon excited states~\cite{White1993,*White2008}.

\begin{figure}
\centering
\includegraphics[clip,trim=0cm 0cm 0cm 6cm,width=\linewidth]{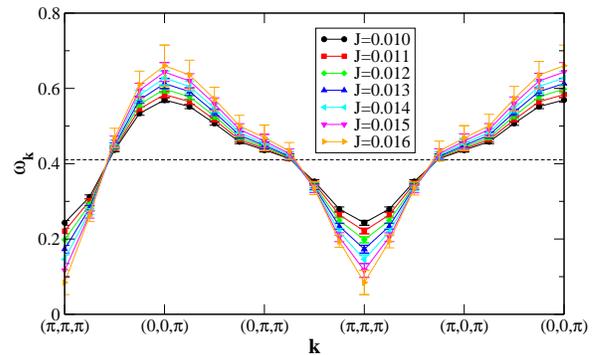}
\caption{(Color online) Dispersion as interchain coupling is increased from Haldane phase towards the N\'{e}el phase. Data shown for length $L=32$ and aspect ratio $R=4$ at inverse temperature $\beta=2L$ at the isotropic point. For comparison, the Haldane gap of a 1D chain\red{\sout{ $0.41050$}}~\cite{White1993} is shown as a dashed line.}
\label{dispersion}
\end{figure}

Plotting the dispersion at $k_z=\pi$ for finite $J$ in Fig.~\ref{dispersion}, it is clear that the Haldane gap closes at ${\bf k}=(\pi,\pi,\pi)$ while the gap at ${\bf k}=(0,0,\pi)$ grows with increasing $J$, implying the presence of interchain correlations. This behavior can be explained by a simple physical argument. In the N\'{e}el state, there exist gapless excitations at ${\bf k}=(0,0,0)$ and ${\bf k}=(\pi,\pi,\pi)$. Thus, starting from the N\'{e}el state and reducing $J$, the finite-$J$ Haldane gap must open up at both ${\bf k}=(0,0,0)$ and ${\bf k}=(\pi,\pi,\pi)$. In general, the Haldane gap must close at the ordering wave vector for any transition to a gapless ordered state. For example, this argument explains the field-driven transition into a canted $XY$-AFM state that has been observed in real Haldane chain systems, such as the orthorhombic material NDMAP~\cite{Zheludev2001,*Zheludev2004}.

\begin{figure}
\centering
\includegraphics[clip,trim=0cm 0cm 0cm 7.10cm,width=\linewidth]{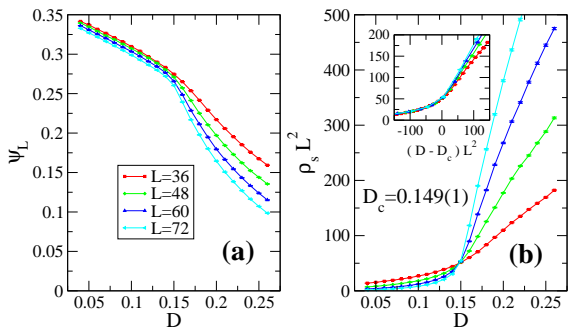}
\caption{(Color online) Finite-size behavior of {\bf (a)} the string order parameter $\Psi_L$ and {\bf (b)} the scaled spin stiffness $\rho_sL^2$ across the Haldane to $XY$-AFM phase boundary with aspect ratio $R=6$, interchain coupling $J=0.01$, and inverse temperature $\beta=L$. The inset shows FSS collapse near the critical point $D_c=0.149(1)$ using mean field critical exponents.}
\label{strings}
\end{figure}

{\em String Order.---}A non-local string order characterizes the Haldane phase in 1D. For Q1D systems, we define string order as the infinite distance limit of string correlations along individual chains~\cite{denNijs1989}
\begin{equation}
C_{SO}\left(i,j\right)=-\left<S_i^z\exp{\left[i\pi\sum_{k=i+1}^{j-1}S_k^z\right]}S_j^z\right>.
\end{equation}
From this we define a finite-size string order parameter $\Psi_L=C_{SO}(0,L/2)$ that scales to zero as $L\rightarrow\infty$ in the absence of string order, or to a finite value when string order is present. This is illustrated in Fig.~\ref{strings}(a), where we show $\Psi_L$ for coupled chains as the single-ion anisotropy $D$ drives the system from the Haldane phase into the $XY$-AFM phase. There is a clear qualitative difference in the finite-size behavior of $\Psi_L$ in the two phases, the boundary of which is determined by FSS of the spin stiffness $\rho_s$ shown in Fig.~\ref{strings}(b). In the $XY$-AFM phase, $\Psi_L$ scales exponentially to zero, while in the Q1D Haldane phase $\Psi_L$ appears to decay algebraically. However, we cannot rule out the possibility that  $\Psi_L$ scales exponentially to zero in the Q1D Haldane phase with correlation length $\xi\gg L$, as predicted for any finite interchain coupling~\cite{Anfuso2007a,*Anfuso2007b}. \blue{Additionally, as string order is ultimately limited to a single dimension, in the following section we introduce a more general correlator for the detection of SPT order.}

\begin{figure}
\centering
\includegraphics[clip,trim=0cm 0cm 0cm 8.1cm,width=\linewidth]{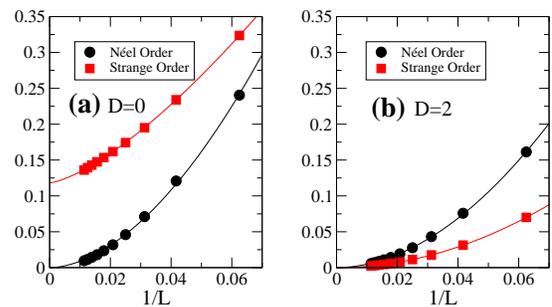}
\caption{(Color online) Finite-size behavior of the N\'{e}el and strange order parameters in {\bf (a)} the Q1D Haldane phase and {\bf (b)} the quantum paramagnetic phase. Results in the ground state limit are obtained using a 2D simulation cell with aspect ratio $R=4$ and interchain coupling $J=0.02$.}
\label{strange}
\end{figure}

{\em Symmetry Protected Topological Order.---}In order to determine the topological nature of the Q1D Haldane phase, we measure the so-called strange correlator ~\cite{You2013}
\begin{equation}
C_{SC}(i,j)=\frac{\langle\Omega|S_i^+S_j^-|\psi\rangle}{\langle\Omega|\psi\rangle}.
\end{equation}
Here, $|\psi\rangle$ is the ground state of ${\cal H}$ and $|\Omega\rangle$ is a trivial product state. \blue{The strange correlator can be thought of as a correlation function at the temporal boundary of the time evolved states $|\Omega\rangle$ and $|\psi\rangle$.} You {\it et al.}~\cite{You2013} have shown that when $|\psi\rangle$ is a non-trivial SPT state, the strange correlator must be long range in 1D and at least \red{\sout{algebraic}}\blue{quasi long range} in 2D\red{\sout{ (note that unlike string correlations, the strange correlator is not confined to 1D)}}. Here, we define a finite-size strange order parameter $\frac{1}{N^2}\sum_{ij} C_{SC}(i,j)$ that \blue{as $L\rightarrow\infty$} scales \blue{exponentially} to zero \red{\sout{as $L\rightarrow\infty$ in the absence of strange order}}\blue{for trivial SPT states, and}\red{\sout{, or}} to a finite value \red{\sout{when strange order is present}}\blue{(or, possibly, algebraically to zero) for non-trivial SPT states}. As shown in Fig.~\ref{strange}(a), the strange order scales to a finite value in the Q1D Haldane phase. This provides direct evidence of non-trivial SPT order in the Q1D Haldane phase. For comparison, Fig.~\ref{strange}(b) shows that the strange order scales to zero in the quantum paramagnetic phase (a trivial SPT state), while in both phases the total staggered magnetization per site (N\'eel order) decays quickly to zero.

{\em Discussion.---}The Haldane phase in 1D has been categorized as a non-trivial SPT state protected by a combination of time-reversal, spin rotation, translation, and spatial inversion symmetries~\cite{Gu2009,*Chen2013,Pollmann2012}. While it seems natural for the Haldane phase to remain a non-trivial SPT state in Q1D, it is known that isotropic integer spin-$S$ HAFM chains have non-trivial SPT states only for $S$ odd, while for $S$ even the ground state can be adiabatically connected to a trivial product state~\cite{Pollmann2012}. This implies that $N$-leg spin-1 HAFM ladders also possess non-trivial SPT characteristics only for $N$ odd, while those with $N$ even are topologically trivial (in agreement with past work on 2-leg~\cite{Todo2001} and 3-leg~\cite{Charrier2010} ladders). Thus, it is not entirely clear what to expect as $N\rightarrow\infty$, which is the Q1D geometry we have considered here (for 2D). Our results provide strong evidence that the \blue{Q1D} Haldane phase of weakly coupled chains in 2D and 3D is indeed a non-trivial SPT state, as determined by probing directly two independent measures of non-local order, viz., the string and strange order parameters.

Interestingly, Matsumoto {\it et al.}~\cite{Matsumoto2001} have demonstrated that the Q1D Haldane phase can be adiabatically connected to a dimer product state. This does not contradict our identification of the Q1D Haldane phase as a non-trivial SPT state, because the geometry they consider is closer to a system of weakly coupled 2-leg ladders than one of weakly coupled chains. In this context, their result is not surprising since 2-leg ladders form trivial SPT states~\cite{Pollmann2012}. \red{\sout{The}}\blue{Actually, the} adiabatic path of Matsumoto {\it et al.}~\cite{Matsumoto2001} breaks the symmetry of spatial inversion about an intrachain bond. We propose this to be a key protecting symmetry of the Q1D Haldane phase.

Further insight is gained by considering the fact that non-trivial SPT states support gapless or degenerate edge states~\cite{Gu2009,*Chen2013}. For example, the Haldane phase in 1D supports degenerate spin-1/2 edge states. If we couple Haldane chains into an $N$-leg ladder, the edge states will form an overall singlet for $N$ even, but retain a degeneracy for $N$ odd due to Kramer's theorem. Thus we understand why even leg ladders form trivial SPT states, while odd leg ladders form non-trivial SPT states. What happens as $N\rightarrow\infty$? In this case, we obtain a gapless spin-1/2 HAFM chain at the edge. However, any bond alternation of the interchain coupling leads to dimer product edge states. Thus, the Q1D Haldane phase is a non-trivial SPT state only for uniformly coupled chains, and spatial inversion symmetry about the intrachain bonds protects against edge state dimerization.

In conclusion, we have accurately determined the ground state phase boundaries of the Haldane phase in weakly coupled spin-1 HAFM chains with uniaxial single-ion anisotropy. The Haldane gap survives up to a critical coupling $J_c$ beyond which it is quenched by magnetic order. By studying \red{\sout{the }}string correlations and \blue{the} strange correlator, we have verified the non-trivial SPT nature of the \blue{Q1D} Haldane phase \red{\sout{in Q1D geometries}}\blue{of the model Hamiltonian~(\ref{hamiltonian})}. We hope our work may motivate further experimental investigations of Q1D spin-1 Heisenberg antiferromagnets.

\begin{acknowledgments}
This research used resources of the National Energy Research Scientific Computing Center, which is supported by the Office of Science of the U.S. Department of Energy under Contract No. DE-AC02-05CH11231. One of us, KW, acknowledge the support of the U.S. National Science Foundation I2CAM International Materials Institute Award, Grant DMR-0844115.
\end{acknowledgments}

\bibliographystyle{apsrev4-1}
\bibliography{haldane-ref}

\begin{thebibliography}{60}%
\makeatletter
\providecommand \@ifxundefined [1]{%
 \@ifx{#1\undefined}
}%
\providecommand \@ifnum [1]{%
 \ifnum #1\expandafter \@firstoftwo
 \else \expandafter \@secondoftwo
 \fi
}%
\providecommand \@ifx [1]{%
 \ifx #1\expandafter \@firstoftwo
 \else \expandafter \@secondoftwo
 \fi
}%
\providecommand \natexlab [1]{#1}%
\providecommand \enquote  [1]{``#1''}%
\providecommand \bibnamefont  [1]{#1}%
\providecommand \bibfnamefont [1]{#1}%
\providecommand \citenamefont [1]{#1}%
\providecommand \href@noop [0]{\@secondoftwo}%
\providecommand \href [0]{\begingroup \@sanitize@url \@href}%
\providecommand \@href[1]{\@@startlink{#1}\@@href}%
\providecommand \@@href[1]{\endgroup#1\@@endlink}%
\providecommand \@sanitize@url [0]{\catcode `\\12\catcode `\$12\catcode
  `\&12\catcode `\#12\catcode `\^12\catcode `\_12\catcode `\%12\relax}%
\providecommand \@@startlink[1]{}%
\providecommand \@@endlink[0]{}%
\providecommand \url  [0]{\begingroup\@sanitize@url \@url }%
\providecommand \@url [1]{\endgroup\@href {#1}{\urlprefix }}%
\providecommand \urlprefix  [0]{URL }%
\providecommand \Eprint [0]{\href }%
\providecommand \doibase [0]{http://dx.doi.org/}%
\providecommand \selectlanguage [0]{\@gobble}%
\providecommand \bibinfo  [0]{\@secondoftwo}%
\providecommand \bibfield  [0]{\@secondoftwo}%
\providecommand \translation [1]{[#1]}%
\providecommand \BibitemOpen [0]{}%
\providecommand \bibitemStop [0]{}%
\providecommand \bibitemNoStop [0]{.\EOS\space}%
\providecommand \EOS [0]{\spacefactor3000\relax}%
\providecommand \BibitemShut  [1]{\csname bibitem#1\endcsname}%
\let\auto@bib@innerbib\@empty
\bibitem [{\citenamefont {Haldane}(1983{\natexlab{a}})}]{Haldane1983a}%
  \BibitemOpen
  \bibfield  {author} {\bibinfo {author} {\bibfnamefont {F.}~\bibnamefont
  {Haldane}},\ }\href {\doibase 10.1016/0375-9601(83)90631-X} {\bibfield
  {journal} {\bibinfo  {journal} {Phys. Lett. A}\ }\textbf {\bibinfo {volume}
  {93}},\ \bibinfo {pages} {464 } (\bibinfo {year}
  {1983}{\natexlab{a}})}\BibitemShut {NoStop}%
\bibitem [{\citenamefont {Haldane}(1983{\natexlab{b}})}]{Haldane1983b}%
  \BibitemOpen
  \bibfield  {author} {\bibinfo {author} {\bibfnamefont {F.~D.~M.}\
  \bibnamefont {Haldane}},\ }\href {\doibase 10.1103/PhysRevLett.50.1153}
  {\bibfield  {journal} {\bibinfo  {journal} {Phys. Rev. Lett.}\ }\textbf
  {\bibinfo {volume} {50}},\ \bibinfo {pages} {1153} (\bibinfo {year}
  {1983}{\natexlab{b}})}\BibitemShut {NoStop}%
\bibitem [{\citenamefont {Sakai}\ and\ \citenamefont
  {Takahashi}(1989)}]{Sakai1989}%
  \BibitemOpen
  \bibfield  {author} {\bibinfo {author} {\bibfnamefont {T.}~\bibnamefont
  {Sakai}}\ and\ \bibinfo {author} {\bibfnamefont {M.}~\bibnamefont
  {Takahashi}},\ }\href {\doibase 10.1143/JPSJ.58.3131} {\bibfield  {journal}
  {\bibinfo  {journal} {J. Phys. Soc. Jpn.}\ }\textbf {\bibinfo {volume}
  {58}},\ \bibinfo {pages} {3131} (\bibinfo {year} {1989})}\BibitemShut
  {NoStop}%
\bibitem [{\citenamefont {Sakai}\ and\ \citenamefont
  {Takahashi}(1990)}]{Sakai1990}%
  \BibitemOpen
  \bibfield  {author} {\bibinfo {author} {\bibfnamefont {T.}~\bibnamefont
  {Sakai}}\ and\ \bibinfo {author} {\bibfnamefont {M.}~\bibnamefont
  {Takahashi}},\ }\href {\doibase 10.1103/PhysRevB.42.4537} {\bibfield
  {journal} {\bibinfo  {journal} {Phys. Rev. B}\ }\textbf {\bibinfo {volume}
  {42}},\ \bibinfo {pages} {4537} (\bibinfo {year} {1990})}\BibitemShut
  {NoStop}%
\bibitem [{\citenamefont {Botet}\ \emph {et~al.}(1983)\citenamefont {Botet},
  \citenamefont {Jullien},\ and\ \citenamefont {Kolb}}]{Botet1983}%
  \BibitemOpen
  \bibfield  {author} {\bibinfo {author} {\bibfnamefont {R.}~\bibnamefont
  {Botet}}, \bibinfo {author} {\bibfnamefont {R.}~\bibnamefont {Jullien}}, \
  and\ \bibinfo {author} {\bibfnamefont {M.}~\bibnamefont {Kolb}},\ }\href
  {\doibase 10.1103/PhysRevB.28.3914} {\bibfield  {journal} {\bibinfo
  {journal} {Phys. Rev. B}\ }\textbf {\bibinfo {volume} {28}},\ \bibinfo
  {pages} {3914} (\bibinfo {year} {1983})}\BibitemShut {NoStop}%
\bibitem [{\citenamefont {Golinelli}\ \emph {et~al.}(1992)\citenamefont
  {Golinelli}, \citenamefont {Jolicoeur},\ and\ \citenamefont
  {Lacaze}}]{Golinelli1992}%
  \BibitemOpen
  \bibfield  {author} {\bibinfo {author} {\bibfnamefont {O.}~\bibnamefont
  {Golinelli}}, \bibinfo {author} {\bibfnamefont {T.}~\bibnamefont
  {Jolicoeur}}, \ and\ \bibinfo {author} {\bibfnamefont {R.}~\bibnamefont
  {Lacaze}},\ }\href {\doibase 10.1103/PhysRevB.45.9798} {\bibfield  {journal}
  {\bibinfo  {journal} {Phys. Rev. B}\ }\textbf {\bibinfo {volume} {45}},\
  \bibinfo {pages} {9798} (\bibinfo {year} {1992})}\BibitemShut {NoStop}%
\bibitem [{\citenamefont {White}\ and\ \citenamefont {Huse}(1993)}]{White1993}%
  \BibitemOpen
  \bibfield  {author} {\bibinfo {author} {\bibfnamefont {S.~R.}\ \bibnamefont
  {White}}\ and\ \bibinfo {author} {\bibfnamefont {D.~A.}\ \bibnamefont
  {Huse}},\ }\href {\doibase 10.1103/PhysRevB.48.3844} {\bibfield  {journal}
  {\bibinfo  {journal} {Phys. Rev. B}\ }\textbf {\bibinfo {volume} {48}},\
  \bibinfo {pages} {3844} (\bibinfo {year} {1993})}\BibitemShut {NoStop}%
\bibitem [{\citenamefont {White}\ and\ \citenamefont
  {Affleck}(2008)}]{White2008}%
  \BibitemOpen
  \bibfield  {author} {\bibinfo {author} {\bibfnamefont {S.~R.}\ \bibnamefont
  {White}}\ and\ \bibinfo {author} {\bibfnamefont {I.}~\bibnamefont
  {Affleck}},\ }\href {\doibase 10.1103/PhysRevB.77.134437} {\bibfield
  {journal} {\bibinfo  {journal} {Phys. Rev. B}\ }\textbf {\bibinfo {volume}
  {77}},\ \bibinfo {pages} {134437} (\bibinfo {year} {2008})}\BibitemShut
  {NoStop}%
\bibitem [{\citenamefont {Hu}\ \emph {et~al.}(2011)\citenamefont {Hu},
  \citenamefont {Normand}, \citenamefont {Wang},\ and\ \citenamefont
  {Yu}}]{Hu2011}%
  \BibitemOpen
  \bibfield  {author} {\bibinfo {author} {\bibfnamefont {S.}~\bibnamefont
  {Hu}}, \bibinfo {author} {\bibfnamefont {B.}~\bibnamefont {Normand}},
  \bibinfo {author} {\bibfnamefont {X.}~\bibnamefont {Wang}}, \ and\ \bibinfo
  {author} {\bibfnamefont {L.}~\bibnamefont {Yu}},\ }\href {\doibase
  10.1103/PhysRevB.84.220402} {\bibfield  {journal} {\bibinfo  {journal} {Phys.
  Rev. B}\ }\textbf {\bibinfo {volume} {84}},\ \bibinfo {pages} {220402}
  (\bibinfo {year} {2011})}\BibitemShut {NoStop}%
\bibitem [{\citenamefont {Moukouri}\ and\ \citenamefont
  {Eidelstein}(2011)}]{Moukouri2011}%
  \BibitemOpen
  \bibfield  {author} {\bibinfo {author} {\bibfnamefont {S.}~\bibnamefont
  {Moukouri}}\ and\ \bibinfo {author} {\bibfnamefont {E.}~\bibnamefont
  {Eidelstein}},\ }\href {\doibase 10.1103/PhysRevB.84.193103} {\bibfield
  {journal} {\bibinfo  {journal} {Phys. Rev. B}\ }\textbf {\bibinfo {volume}
  {84}},\ \bibinfo {pages} {193103} (\bibinfo {year} {2011})}\BibitemShut
  {NoStop}%
\bibitem [{\citenamefont {Moukouri}\ and\ \citenamefont
  {Eidelstein}(2012)}]{Moukouri2012}%
  \BibitemOpen
  \bibfield  {author} {\bibinfo {author} {\bibfnamefont {S.}~\bibnamefont
  {Moukouri}}\ and\ \bibinfo {author} {\bibfnamefont {E.}~\bibnamefont
  {Eidelstein}},\ }\href {\doibase 10.1103/PhysRevB.86.155112} {\bibfield
  {journal} {\bibinfo  {journal} {Phys. Rev. B}\ }\textbf {\bibinfo {volume}
  {86}},\ \bibinfo {pages} {155112} (\bibinfo {year} {2012})}\BibitemShut
  {NoStop}%
\bibitem [{\citenamefont {Albuquerque}\ \emph {et~al.}(2009)\citenamefont
  {Albuquerque}, \citenamefont {Hamer},\ and\ \citenamefont
  {Oitmaa}}]{Albuquerque2009}%
  \BibitemOpen
  \bibfield  {author} {\bibinfo {author} {\bibfnamefont {A.~F.}\ \bibnamefont
  {Albuquerque}}, \bibinfo {author} {\bibfnamefont {C.~J.}\ \bibnamefont
  {Hamer}}, \ and\ \bibinfo {author} {\bibfnamefont {J.}~\bibnamefont
  {Oitmaa}},\ }\href {\doibase 10.1103/PhysRevB.79.054412} {\bibfield
  {journal} {\bibinfo  {journal} {Phys. Rev. B}\ }\textbf {\bibinfo {volume}
  {79}},\ \bibinfo {pages} {054412} (\bibinfo {year} {2009})}\BibitemShut
  {NoStop}%
\bibitem [{\citenamefont {Kim}\ and\ \citenamefont
  {Birgeneau}(2000)}]{Kim2000}%
  \BibitemOpen
  \bibfield  {author} {\bibinfo {author} {\bibfnamefont {Y.~J.}\ \bibnamefont
  {Kim}}\ and\ \bibinfo {author} {\bibfnamefont {R.~J.}\ \bibnamefont
  {Birgeneau}},\ }\href {\doibase 10.1103/PhysRevB.62.6378} {\bibfield
  {journal} {\bibinfo  {journal} {Phys. Rev. B}\ }\textbf {\bibinfo {volume}
  {62}},\ \bibinfo {pages} {6378} (\bibinfo {year} {2000})}\BibitemShut
  {NoStop}%
\bibitem [{\citenamefont {Matsumoto}\ \emph {et~al.}(2001)\citenamefont
  {Matsumoto}, \citenamefont {Yasuda}, \citenamefont {Todo},\ and\
  \citenamefont {Takayama}}]{Matsumoto2001}%
  \BibitemOpen
  \bibfield  {author} {\bibinfo {author} {\bibfnamefont {M.}~\bibnamefont
  {Matsumoto}}, \bibinfo {author} {\bibfnamefont {C.}~\bibnamefont {Yasuda}},
  \bibinfo {author} {\bibfnamefont {S.}~\bibnamefont {Todo}}, \ and\ \bibinfo
  {author} {\bibfnamefont {H.}~\bibnamefont {Takayama}},\ }\href {\doibase
  10.1103/PhysRevB.65.014407} {\bibfield  {journal} {\bibinfo  {journal} {Phys.
  Rev. B}\ }\textbf {\bibinfo {volume} {65}},\ \bibinfo {pages} {014407}
  (\bibinfo {year} {2001})}\BibitemShut {NoStop}%
\bibitem [{\citenamefont {Wierschem}\ and\ \citenamefont
  {Sengupta}(2012)}]{Wierschem2012}%
  \BibitemOpen
  \bibfield  {author} {\bibinfo {author} {\bibfnamefont {K.}~\bibnamefont
  {Wierschem}}\ and\ \bibinfo {author} {\bibfnamefont {P.}~\bibnamefont
  {Sengupta}},\ }\href {\doibase 10.1088/1742-6596/400/3/032112} {\bibfield
  {journal} {\bibinfo  {journal} {J. Phys.: Conf. Ser.}\ }\textbf {\bibinfo
  {volume} {400}},\ \bibinfo {pages} {032112} (\bibinfo {year}
  {2012})}\BibitemShut {NoStop}%
\bibitem [{\citenamefont {Wierschem}\ and\ \citenamefont
  {Sengupta}(2013)}]{Wierschem2013}%
  \BibitemOpen
  \bibfield  {author} {\bibinfo {author} {\bibfnamefont {K.}~\bibnamefont
  {Wierschem}}\ and\ \bibinfo {author} {\bibfnamefont {P.}~\bibnamefont
  {Sengupta}},\ }\href@noop {} {} (\bibinfo {year} {2013}),\ \bibinfo {note}
  {arXiv:1310:0130}\BibitemShut {NoStop}%
\bibitem [{\citenamefont {Mutka}\ \emph {et~al.}(1991)\citenamefont {Mutka},
  \citenamefont {Payen}, \citenamefont {Molini\'e}, \citenamefont {Soubeyroux},
  \citenamefont {Colombet},\ and\ \citenamefont {Taylor}}]{Mutka1991}%
  \BibitemOpen
  \bibfield  {author} {\bibinfo {author} {\bibfnamefont {H.}~\bibnamefont
  {Mutka}}, \bibinfo {author} {\bibfnamefont {C.}~\bibnamefont {Payen}},
  \bibinfo {author} {\bibfnamefont {P.}~\bibnamefont {Molini\'e}}, \bibinfo
  {author} {\bibfnamefont {J.~L.}\ \bibnamefont {Soubeyroux}}, \bibinfo
  {author} {\bibfnamefont {P.}~\bibnamefont {Colombet}}, \ and\ \bibinfo
  {author} {\bibfnamefont {A.~D.}\ \bibnamefont {Taylor}},\ }\href {\doibase
  10.1103/PhysRevLett.67.497} {\bibfield  {journal} {\bibinfo  {journal} {Phys.
  Rev. Lett.}\ }\textbf {\bibinfo {volume} {67}},\ \bibinfo {pages} {497}
  (\bibinfo {year} {1991})}\BibitemShut {NoStop}%
\bibitem [{\citenamefont {Asano}\ \emph {et~al.}(1994)\citenamefont {Asano},
  \citenamefont {Ajiro}, \citenamefont {Mekata}, \citenamefont {Yamazaki},
  \citenamefont {Hosoito}, \citenamefont {Shinjo},\ and\ \citenamefont
  {Kikuchi}}]{Asano1994}%
  \BibitemOpen
  \bibfield  {author} {\bibinfo {author} {\bibfnamefont {T.}~\bibnamefont
  {Asano}}, \bibinfo {author} {\bibfnamefont {Y.}~\bibnamefont {Ajiro}},
  \bibinfo {author} {\bibfnamefont {M.}~\bibnamefont {Mekata}}, \bibinfo
  {author} {\bibfnamefont {H.}~\bibnamefont {Yamazaki}}, \bibinfo {author}
  {\bibfnamefont {N.}~\bibnamefont {Hosoito}}, \bibinfo {author} {\bibfnamefont
  {T.}~\bibnamefont {Shinjo}}, \ and\ \bibinfo {author} {\bibfnamefont
  {H.}~\bibnamefont {Kikuchi}},\ }\href {\doibase
  http://dx.doi.org/10.1016/0038-1098(94)90944-X} {\bibfield  {journal}
  {\bibinfo  {journal} {Solid State Commun.}\ }\textbf {\bibinfo {volume}
  {90}},\ \bibinfo {pages} {125 } (\bibinfo {year} {1994})}\BibitemShut
  {NoStop}%
\bibitem [{\citenamefont {Takigawa}\ \emph {et~al.}(1995)\citenamefont
  {Takigawa}, \citenamefont {Asano}, \citenamefont {Ajiro},\ and\ \citenamefont
  {Mekata}}]{Takigawa1995}%
  \BibitemOpen
  \bibfield  {author} {\bibinfo {author} {\bibfnamefont {M.}~\bibnamefont
  {Takigawa}}, \bibinfo {author} {\bibfnamefont {T.}~\bibnamefont {Asano}},
  \bibinfo {author} {\bibfnamefont {Y.}~\bibnamefont {Ajiro}}, \ and\ \bibinfo
  {author} {\bibfnamefont {M.}~\bibnamefont {Mekata}},\ }\href {\doibase
  10.1103/PhysRevB.52.R13087} {\bibfield  {journal} {\bibinfo  {journal} {Phys.
  Rev. B}\ }\textbf {\bibinfo {volume} {52}},\ \bibinfo {pages} {R13087}
  (\bibinfo {year} {1995})}\BibitemShut {NoStop}%
\bibitem [{\citenamefont {Takigawa}\ \emph {et~al.}(1996)\citenamefont
  {Takigawa}, \citenamefont {Asano}, \citenamefont {Ajiro}, \citenamefont
  {Mekata},\ and\ \citenamefont {Uemura}}]{Takigawa1996}%
  \BibitemOpen
  \bibfield  {author} {\bibinfo {author} {\bibfnamefont {M.}~\bibnamefont
  {Takigawa}}, \bibinfo {author} {\bibfnamefont {T.}~\bibnamefont {Asano}},
  \bibinfo {author} {\bibfnamefont {Y.}~\bibnamefont {Ajiro}}, \bibinfo
  {author} {\bibfnamefont {M.}~\bibnamefont {Mekata}}, \ and\ \bibinfo {author}
  {\bibfnamefont {Y.~J.}\ \bibnamefont {Uemura}},\ }\href {\doibase
  10.1103/PhysRevLett.76.2173} {\bibfield  {journal} {\bibinfo  {journal}
  {Phys. Rev. Lett.}\ }\textbf {\bibinfo {volume} {76}},\ \bibinfo {pages}
  {2173} (\bibinfo {year} {1996})}\BibitemShut {NoStop}%
\bibitem [{\citenamefont {Honda}\ \emph {et~al.}(2001)\citenamefont {Honda},
  \citenamefont {Katsumata}, \citenamefont {Nishiyama},\ and\ \citenamefont
  {Harada}}]{Honda2001}%
  \BibitemOpen
  \bibfield  {author} {\bibinfo {author} {\bibfnamefont {Z.}~\bibnamefont
  {Honda}}, \bibinfo {author} {\bibfnamefont {K.}~\bibnamefont {Katsumata}},
  \bibinfo {author} {\bibfnamefont {Y.}~\bibnamefont {Nishiyama}}, \ and\
  \bibinfo {author} {\bibfnamefont {I.}~\bibnamefont {Harada}},\ }\href
  {\doibase 10.1103/PhysRevB.63.064420} {\bibfield  {journal} {\bibinfo
  {journal} {Phys. Rev. B}\ }\textbf {\bibinfo {volume} {63}},\ \bibinfo
  {pages} {064420} (\bibinfo {year} {2001})}\BibitemShut {NoStop}%
\bibitem [{\citenamefont {Renard}\ \emph {et~al.}(1987)\citenamefont {Renard},
  \citenamefont {Verdaguer}, \citenamefont {Regnault}, \citenamefont
  {Erkelens}, \citenamefont {Rossat-Mignod},\ and\ \citenamefont
  {Stirling}}]{Renard1987}%
  \BibitemOpen
  \bibfield  {author} {\bibinfo {author} {\bibfnamefont {J.~P.}\ \bibnamefont
  {Renard}}, \bibinfo {author} {\bibfnamefont {M.}~\bibnamefont {Verdaguer}},
  \bibinfo {author} {\bibfnamefont {L.~P.}\ \bibnamefont {Regnault}}, \bibinfo
  {author} {\bibfnamefont {W.~A.~C.}\ \bibnamefont {Erkelens}}, \bibinfo
  {author} {\bibfnamefont {J.}~\bibnamefont {Rossat-Mignod}}, \ and\ \bibinfo
  {author} {\bibfnamefont {W.~G.}\ \bibnamefont {Stirling}},\ }\href@noop {}
  {\bibfield  {journal} {\bibinfo  {journal} {Europhys. Lett.}\ }\textbf
  {\bibinfo {volume} {3}},\ \bibinfo {pages} {945} (\bibinfo {year}
  {1987})}\BibitemShut {NoStop}%
\bibitem [{\citenamefont {Renard}\ \emph {et~al.}(1988)\citenamefont {Renard},
  \citenamefont {Verdaguer}, \citenamefont {Regnault}, \citenamefont
  {Erkelens}, \citenamefont {Rossat-Mignod}, \citenamefont {Ribas},
  \citenamefont {Stirling},\ and\ \citenamefont {Vettier}}]{Renard1988}%
  \BibitemOpen
  \bibfield  {author} {\bibinfo {author} {\bibfnamefont {J.~P.}\ \bibnamefont
  {Renard}}, \bibinfo {author} {\bibfnamefont {M.}~\bibnamefont {Verdaguer}},
  \bibinfo {author} {\bibfnamefont {L.~P.}\ \bibnamefont {Regnault}}, \bibinfo
  {author} {\bibfnamefont {W.~A.~C.}\ \bibnamefont {Erkelens}}, \bibinfo
  {author} {\bibfnamefont {J.}~\bibnamefont {Rossat-Mignod}}, \bibinfo {author}
  {\bibfnamefont {J.}~\bibnamefont {Ribas}}, \bibinfo {author} {\bibfnamefont
  {W.~G.}\ \bibnamefont {Stirling}}, \ and\ \bibinfo {author} {\bibfnamefont
  {C.}~\bibnamefont {Vettier}},\ }\href {\doibase
  http://dx.doi.org/10.1063/1.340736} {\bibfield  {journal} {\bibinfo
  {journal} {J. Appl. Phys.}\ }\textbf {\bibinfo {volume} {63}},\ \bibinfo
  {pages} {3538} (\bibinfo {year} {1988})}\BibitemShut {NoStop}%
\bibitem [{\citenamefont {Regnault}\ \emph {et~al.}(1994)\citenamefont
  {Regnault}, \citenamefont {Zaliznyak}, \citenamefont {Renard},\ and\
  \citenamefont {Vettier}}]{Regnault1994}%
  \BibitemOpen
  \bibfield  {author} {\bibinfo {author} {\bibfnamefont {L.~P.}\ \bibnamefont
  {Regnault}}, \bibinfo {author} {\bibfnamefont {I.}~\bibnamefont {Zaliznyak}},
  \bibinfo {author} {\bibfnamefont {J.~P.}\ \bibnamefont {Renard}}, \ and\
  \bibinfo {author} {\bibfnamefont {C.}~\bibnamefont {Vettier}},\ }\href
  {\doibase 10.1103/PhysRevB.50.9174} {\bibfield  {journal} {\bibinfo
  {journal} {Phys. Rev. B}\ }\textbf {\bibinfo {volume} {50}},\ \bibinfo
  {pages} {9174} (\bibinfo {year} {1994})}\BibitemShut {NoStop}%
\bibitem [{\citenamefont {Zaliznyak}\ \emph {et~al.}(1998)\citenamefont
  {Zaliznyak}, \citenamefont {Dender}, \citenamefont {Broholm},\ and\
  \citenamefont {Reich}}]{Zaliznyak1998}%
  \BibitemOpen
  \bibfield  {author} {\bibinfo {author} {\bibfnamefont {I.~A.}\ \bibnamefont
  {Zaliznyak}}, \bibinfo {author} {\bibfnamefont {D.~C.}\ \bibnamefont
  {Dender}}, \bibinfo {author} {\bibfnamefont {C.}~\bibnamefont {Broholm}}, \
  and\ \bibinfo {author} {\bibfnamefont {D.~H.}\ \bibnamefont {Reich}},\ }\href
  {\doibase 10.1103/PhysRevB.57.5200} {\bibfield  {journal} {\bibinfo
  {journal} {Phys. Rev. B}\ }\textbf {\bibinfo {volume} {57}},\ \bibinfo
  {pages} {5200} (\bibinfo {year} {1998})}\BibitemShut {NoStop}%
\bibitem [{\citenamefont {Uchiyama}\ \emph {et~al.}(1999)\citenamefont
  {Uchiyama}, \citenamefont {Sasago}, \citenamefont {Tsukada}, \citenamefont
  {Uchinokura}, \citenamefont {Zheludev}, \citenamefont {Hayashi},
  \citenamefont {Miura},\ and\ \citenamefont {B\"oni}}]{Uchiyama1999}%
  \BibitemOpen
  \bibfield  {author} {\bibinfo {author} {\bibfnamefont {Y.}~\bibnamefont
  {Uchiyama}}, \bibinfo {author} {\bibfnamefont {Y.}~\bibnamefont {Sasago}},
  \bibinfo {author} {\bibfnamefont {I.}~\bibnamefont {Tsukada}}, \bibinfo
  {author} {\bibfnamefont {K.}~\bibnamefont {Uchinokura}}, \bibinfo {author}
  {\bibfnamefont {A.}~\bibnamefont {Zheludev}}, \bibinfo {author}
  {\bibfnamefont {T.}~\bibnamefont {Hayashi}}, \bibinfo {author} {\bibfnamefont
  {N.}~\bibnamefont {Miura}}, \ and\ \bibinfo {author} {\bibfnamefont
  {P.}~\bibnamefont {B\"oni}},\ }\href {\doibase 10.1103/PhysRevLett.83.632}
  {\bibfield  {journal} {\bibinfo  {journal} {Phys. Rev. Lett.}\ }\textbf
  {\bibinfo {volume} {83}},\ \bibinfo {pages} {632} (\bibinfo {year}
  {1999})}\BibitemShut {NoStop}%
\bibitem [{\citenamefont {Zheludev}\ \emph {et~al.}(2000)\citenamefont
  {Zheludev}, \citenamefont {Masuda}, \citenamefont {Tsukada}, \citenamefont
  {Uchiyama}, \citenamefont {Uchinokura}, \citenamefont {B\"oni},\ and\
  \citenamefont {Lee}}]{Zheludev2000}%
  \BibitemOpen
  \bibfield  {author} {\bibinfo {author} {\bibfnamefont {A.}~\bibnamefont
  {Zheludev}}, \bibinfo {author} {\bibfnamefont {T.}~\bibnamefont {Masuda}},
  \bibinfo {author} {\bibfnamefont {I.}~\bibnamefont {Tsukada}}, \bibinfo
  {author} {\bibfnamefont {Y.}~\bibnamefont {Uchiyama}}, \bibinfo {author}
  {\bibfnamefont {K.}~\bibnamefont {Uchinokura}}, \bibinfo {author}
  {\bibfnamefont {P.}~\bibnamefont {B\"oni}}, \ and\ \bibinfo {author}
  {\bibfnamefont {S.-H.}\ \bibnamefont {Lee}},\ }\href {\doibase
  10.1103/PhysRevB.62.8921} {\bibfield  {journal} {\bibinfo  {journal} {Phys.
  Rev. B}\ }\textbf {\bibinfo {volume} {62}},\ \bibinfo {pages} {8921}
  (\bibinfo {year} {2000})}\BibitemShut {NoStop}%
\bibitem [{\citenamefont {Pahari}\ \emph {et~al.}(2006)\citenamefont {Pahari},
  \citenamefont {Ghoshray}, \citenamefont {Sarkar}, \citenamefont
  {Bandyopadhyay},\ and\ \citenamefont {Ghoshray}}]{Pahari2006}%
  \BibitemOpen
  \bibfield  {author} {\bibinfo {author} {\bibfnamefont {B.}~\bibnamefont
  {Pahari}}, \bibinfo {author} {\bibfnamefont {K.}~\bibnamefont {Ghoshray}},
  \bibinfo {author} {\bibfnamefont {R.}~\bibnamefont {Sarkar}}, \bibinfo
  {author} {\bibfnamefont {B.}~\bibnamefont {Bandyopadhyay}}, \ and\ \bibinfo
  {author} {\bibfnamefont {A.}~\bibnamefont {Ghoshray}},\ }\href {\doibase
  10.1103/PhysRevB.73.012407} {\bibfield  {journal} {\bibinfo  {journal} {Phys.
  Rev. B}\ }\textbf {\bibinfo {volume} {73}},\ \bibinfo {pages} {012407}
  (\bibinfo {year} {2006})}\BibitemShut {NoStop}%
\bibitem [{\citenamefont {Gadet}\ \emph {et~al.}(1991)\citenamefont {Gadet},
  \citenamefont {Verdaguer}, \citenamefont {Briois}, \citenamefont {Gleizes},
  \citenamefont {Renard}, \citenamefont {Beauvillain}, \citenamefont
  {Chappert}, \citenamefont {Goto}, \citenamefont {Le~Dang},\ and\
  \citenamefont {Veillet}}]{Gadet1991}%
  \BibitemOpen
  \bibfield  {author} {\bibinfo {author} {\bibfnamefont {V.}~\bibnamefont
  {Gadet}}, \bibinfo {author} {\bibfnamefont {M.}~\bibnamefont {Verdaguer}},
  \bibinfo {author} {\bibfnamefont {V.}~\bibnamefont {Briois}}, \bibinfo
  {author} {\bibfnamefont {A.}~\bibnamefont {Gleizes}}, \bibinfo {author}
  {\bibfnamefont {J.~P.}\ \bibnamefont {Renard}}, \bibinfo {author}
  {\bibfnamefont {P.}~\bibnamefont {Beauvillain}}, \bibinfo {author}
  {\bibfnamefont {C.}~\bibnamefont {Chappert}}, \bibinfo {author}
  {\bibfnamefont {T.}~\bibnamefont {Goto}}, \bibinfo {author} {\bibfnamefont
  {K.}~\bibnamefont {Le~Dang}}, \ and\ \bibinfo {author} {\bibfnamefont
  {P.}~\bibnamefont {Veillet}},\ }\href {\doibase 10.1103/PhysRevB.44.705}
  {\bibfield  {journal} {\bibinfo  {journal} {Phys. Rev. B}\ }\textbf {\bibinfo
  {volume} {44}},\ \bibinfo {pages} {705} (\bibinfo {year} {1991})}\BibitemShut
  {NoStop}%
\bibitem [{\citenamefont {Darriet}\ and\ \citenamefont
  {Regnault}(1993)}]{Darriet1993}%
  \BibitemOpen
  \bibfield  {author} {\bibinfo {author} {\bibfnamefont {J.}~\bibnamefont
  {Darriet}}\ and\ \bibinfo {author} {\bibfnamefont {L.}~\bibnamefont
  {Regnault}},\ }\href {\doibase
  http://dx.doi.org/10.1016/0038-1098(93)90455-V} {\bibfield  {journal}
  {\bibinfo  {journal} {Solid State Commun.}\ }\textbf {\bibinfo {volume}
  {86}},\ \bibinfo {pages} {409 } (\bibinfo {year} {1993})}\BibitemShut
  {NoStop}%
\bibitem [{\citenamefont {Xu}\ \emph {et~al.}(1996)\citenamefont {Xu},
  \citenamefont {DiTusa}, \citenamefont {Ito}, \citenamefont {Oka},
  \citenamefont {Takagi}, \citenamefont {Broholm},\ and\ \citenamefont
  {Aeppli}}]{Xu1996}%
  \BibitemOpen
  \bibfield  {author} {\bibinfo {author} {\bibfnamefont {G.}~\bibnamefont
  {Xu}}, \bibinfo {author} {\bibfnamefont {J.~F.}\ \bibnamefont {DiTusa}},
  \bibinfo {author} {\bibfnamefont {T.}~\bibnamefont {Ito}}, \bibinfo {author}
  {\bibfnamefont {K.}~\bibnamefont {Oka}}, \bibinfo {author} {\bibfnamefont
  {H.}~\bibnamefont {Takagi}}, \bibinfo {author} {\bibfnamefont
  {C.}~\bibnamefont {Broholm}}, \ and\ \bibinfo {author} {\bibfnamefont
  {G.}~\bibnamefont {Aeppli}},\ }\href {\doibase 10.1103/PhysRevB.54.R6827}
  {\bibfield  {journal} {\bibinfo  {journal} {Phys. Rev. B}\ }\textbf {\bibinfo
  {volume} {54}},\ \bibinfo {pages} {R6827} (\bibinfo {year}
  {1996})}\BibitemShut {NoStop}%
\bibitem [{\citenamefont {R\"uegg}\ \emph {et~al.}(2004)\citenamefont
  {R\"uegg}, \citenamefont {Furrer}, \citenamefont {Sheptyakov}, \citenamefont
  {Str\"assle}, \citenamefont {Kr\"amer}, \citenamefont {G\"udel},\ and\
  \citenamefont {M\'el\'esi}}]{Ruegg2004}%
  \BibitemOpen
  \bibfield  {author} {\bibinfo {author} {\bibfnamefont {C.}~\bibnamefont
  {R\"uegg}}, \bibinfo {author} {\bibfnamefont {A.}~\bibnamefont {Furrer}},
  \bibinfo {author} {\bibfnamefont {D.}~\bibnamefont {Sheptyakov}}, \bibinfo
  {author} {\bibfnamefont {T.}~\bibnamefont {Str\"assle}}, \bibinfo {author}
  {\bibfnamefont {K.~W.}\ \bibnamefont {Kr\"amer}}, \bibinfo {author}
  {\bibfnamefont {H.-U.}\ \bibnamefont {G\"udel}}, \ and\ \bibinfo {author}
  {\bibfnamefont {L.}~\bibnamefont {M\'el\'esi}},\ }\href {\doibase
  10.1103/PhysRevLett.93.257201} {\bibfield  {journal} {\bibinfo  {journal}
  {Phys. Rev. Lett.}\ }\textbf {\bibinfo {volume} {93}},\ \bibinfo {pages}
  {257201} (\bibinfo {year} {2004})}\BibitemShut {NoStop}%
\bibitem [{\citenamefont {Zapf}\ \emph {et~al.}(2006)\citenamefont {Zapf},
  \citenamefont {Zocco}, \citenamefont {Hansen}, \citenamefont {Jaime},
  \citenamefont {Harrison}, \citenamefont {Batista}, \citenamefont
  {Kenzelmann}, \citenamefont {Niedermayer}, \citenamefont {Lacerda},\ and\
  \citenamefont {Paduan-Filho}}]{Zapf2006}%
  \BibitemOpen
  \bibfield  {author} {\bibinfo {author} {\bibfnamefont {V.~S.}\ \bibnamefont
  {Zapf}}, \bibinfo {author} {\bibfnamefont {D.}~\bibnamefont {Zocco}},
  \bibinfo {author} {\bibfnamefont {B.~R.}\ \bibnamefont {Hansen}}, \bibinfo
  {author} {\bibfnamefont {M.}~\bibnamefont {Jaime}}, \bibinfo {author}
  {\bibfnamefont {N.}~\bibnamefont {Harrison}}, \bibinfo {author}
  {\bibfnamefont {C.~D.}\ \bibnamefont {Batista}}, \bibinfo {author}
  {\bibfnamefont {M.}~\bibnamefont {Kenzelmann}}, \bibinfo {author}
  {\bibfnamefont {C.}~\bibnamefont {Niedermayer}}, \bibinfo {author}
  {\bibfnamefont {A.}~\bibnamefont {Lacerda}}, \ and\ \bibinfo {author}
  {\bibfnamefont {A.}~\bibnamefont {Paduan-Filho}},\ }\href {\doibase
  10.1103/PhysRevLett.96.077204} {\bibfield  {journal} {\bibinfo  {journal}
  {Phys. Rev. Lett.}\ }\textbf {\bibinfo {volume} {96}},\ \bibinfo {pages}
  {077204} (\bibinfo {year} {2006})}\BibitemShut {NoStop}%
\bibitem [{\citenamefont {Yin}\ \emph {et~al.}(2008)\citenamefont {Yin},
  \citenamefont {Xia}, \citenamefont {Zapf}, \citenamefont {Sullivan},\ and\
  \citenamefont {Paduan-Filho}}]{Yin2008}%
  \BibitemOpen
  \bibfield  {author} {\bibinfo {author} {\bibfnamefont {L.}~\bibnamefont
  {Yin}}, \bibinfo {author} {\bibfnamefont {J.~S.}\ \bibnamefont {Xia}},
  \bibinfo {author} {\bibfnamefont {V.~S.}\ \bibnamefont {Zapf}}, \bibinfo
  {author} {\bibfnamefont {N.~S.}\ \bibnamefont {Sullivan}}, \ and\ \bibinfo
  {author} {\bibfnamefont {A.}~\bibnamefont {Paduan-Filho}},\ }\href {\doibase
  10.1103/PhysRevLett.101.187205} {\bibfield  {journal} {\bibinfo  {journal}
  {Phys. Rev. Lett.}\ }\textbf {\bibinfo {volume} {101}},\ \bibinfo {pages}
  {187205} (\bibinfo {year} {2008})}\BibitemShut {NoStop}%
\bibitem [{\citenamefont {Sengupta}\ and\ \citenamefont
  {Batista}(2007{\natexlab{a}})}]{Sengupta2007a}%
  \BibitemOpen
  \bibfield  {author} {\bibinfo {author} {\bibfnamefont {P.}~\bibnamefont
  {Sengupta}}\ and\ \bibinfo {author} {\bibfnamefont {C.~D.}\ \bibnamefont
  {Batista}},\ }\href {\doibase 10.1103/PhysRevLett.99.217205} {\bibfield
  {journal} {\bibinfo  {journal} {Phys. Rev. Lett.}\ }\textbf {\bibinfo
  {volume} {99}},\ \bibinfo {pages} {217205} (\bibinfo {year}
  {2007}{\natexlab{a}})}\BibitemShut {NoStop}%
\bibitem [{\citenamefont {Sengupta}\ and\ \citenamefont
  {Batista}(2007{\natexlab{b}})}]{Sengupta2007b}%
  \BibitemOpen
  \bibfield  {author} {\bibinfo {author} {\bibfnamefont {P.}~\bibnamefont
  {Sengupta}}\ and\ \bibinfo {author} {\bibfnamefont {C.~D.}\ \bibnamefont
  {Batista}},\ }\href {\doibase 10.1103/PhysRevLett.98.227201} {\bibfield
  {journal} {\bibinfo  {journal} {Phys. Rev. Lett.}\ }\textbf {\bibinfo
  {volume} {98}},\ \bibinfo {pages} {227201} (\bibinfo {year}
  {2007}{\natexlab{b}})}\BibitemShut {NoStop}%
\bibitem [{\citenamefont {Wierschem}\ \emph {et~al.}(2012)\citenamefont
  {Wierschem}, \citenamefont {Kato}, \citenamefont {Nishida}, \citenamefont
  {Batista},\ and\ \citenamefont {Sengupta}}]{Wierschem2012f}%
  \BibitemOpen
  \bibfield  {author} {\bibinfo {author} {\bibfnamefont {K.}~\bibnamefont
  {Wierschem}}, \bibinfo {author} {\bibfnamefont {Y.}~\bibnamefont {Kato}},
  \bibinfo {author} {\bibfnamefont {Y.}~\bibnamefont {Nishida}}, \bibinfo
  {author} {\bibfnamefont {C.~D.}\ \bibnamefont {Batista}}, \ and\ \bibinfo
  {author} {\bibfnamefont {P.}~\bibnamefont {Sengupta}},\ }\href {\doibase
  10.1103/PhysRevB.86.201108} {\bibfield  {journal} {\bibinfo  {journal} {Phys.
  Rev. B}\ }\textbf {\bibinfo {volume} {86}},\ \bibinfo {pages} {201108}
  (\bibinfo {year} {2012})}\BibitemShut {NoStop}%
\bibitem [{\citenamefont {Gu}\ and\ \citenamefont {Wen}(2009)}]{Gu2009}%
  \BibitemOpen
  \bibfield  {author} {\bibinfo {author} {\bibfnamefont {Z.-C.}\ \bibnamefont
  {Gu}}\ and\ \bibinfo {author} {\bibfnamefont {X.-G.}\ \bibnamefont {Wen}},\
  }\href {\doibase 10.1103/PhysRevB.80.155131} {\bibfield  {journal} {\bibinfo
  {journal} {Phys. Rev. B}\ }\textbf {\bibinfo {volume} {80}},\ \bibinfo
  {pages} {155131} (\bibinfo {year} {2009})}\BibitemShut {NoStop}%
\bibitem [{\citenamefont {Chen}\ \emph {et~al.}(2013)\citenamefont {Chen},
  \citenamefont {Gu}, \citenamefont {Liu},\ and\ \citenamefont
  {Wen}}]{Chen2013}%
  \BibitemOpen
  \bibfield  {author} {\bibinfo {author} {\bibfnamefont {X.}~\bibnamefont
  {Chen}}, \bibinfo {author} {\bibfnamefont {Z.-C.}\ \bibnamefont {Gu}},
  \bibinfo {author} {\bibfnamefont {Z.-X.}\ \bibnamefont {Liu}}, \ and\
  \bibinfo {author} {\bibfnamefont {X.-G.}\ \bibnamefont {Wen}},\ }\href
  {\doibase 10.1103/PhysRevB.87.155114} {\bibfield  {journal} {\bibinfo
  {journal} {Phys. Rev. B}\ }\textbf {\bibinfo {volume} {87}},\ \bibinfo
  {pages} {155114} (\bibinfo {year} {2013})}\BibitemShut {NoStop}%
\bibitem [{\citenamefont {Pollmann}\ \emph {et~al.}(2012)\citenamefont
  {Pollmann}, \citenamefont {Berg}, \citenamefont {Turner},\ and\ \citenamefont
  {Oshikawa}}]{Pollmann2012}%
  \BibitemOpen
  \bibfield  {author} {\bibinfo {author} {\bibfnamefont {F.}~\bibnamefont
  {Pollmann}}, \bibinfo {author} {\bibfnamefont {E.}~\bibnamefont {Berg}},
  \bibinfo {author} {\bibfnamefont {A.~M.}\ \bibnamefont {Turner}}, \ and\
  \bibinfo {author} {\bibfnamefont {M.}~\bibnamefont {Oshikawa}},\ }\href
  {\doibase 10.1103/PhysRevB.85.075125} {\bibfield  {journal} {\bibinfo
  {journal} {Phys. Rev. B}\ }\textbf {\bibinfo {volume} {85}},\ \bibinfo
  {pages} {075125} (\bibinfo {year} {2012})}\BibitemShut {NoStop}%
\bibitem [{Note1()}]{Note1}%
  \BibitemOpen
  \bibinfo {note} {We define the Q1D Haldane phase as the gapped symmetric
  ground state of the model Hamiltonian~(\ref {hamiltonian}) with $J\not =0$
  that is adiabatically connected to the Haldane phase in 1D as $J\rightarrow
  0$}\BibitemShut {NoStop}%
\bibitem [{\citenamefont {Sandvik}(1999)}]{Sandvik1999}%
  \BibitemOpen
  \bibfield  {author} {\bibinfo {author} {\bibfnamefont {A.~W.}\ \bibnamefont
  {Sandvik}},\ }\href {\doibase 10.1103/PhysRevLett.83.3069} {\bibfield
  {journal} {\bibinfo  {journal} {Phys. Rev. Lett.}\ }\textbf {\bibinfo
  {volume} {83}},\ \bibinfo {pages} {3069} (\bibinfo {year}
  {1999})}\BibitemShut {NoStop}%
\bibitem [{\citenamefont {Sandvik}\ and\ \citenamefont
  {Kurkij\"arvi}(1991)}]{Sandvik1991}%
  \BibitemOpen
  \bibfield  {author} {\bibinfo {author} {\bibfnamefont {A.~W.}\ \bibnamefont
  {Sandvik}}\ and\ \bibinfo {author} {\bibfnamefont {J.}~\bibnamefont
  {Kurkij\"arvi}},\ }\href {\doibase 10.1103/PhysRevB.43.5950} {\bibfield
  {journal} {\bibinfo  {journal} {Phys. Rev. B}\ }\textbf {\bibinfo {volume}
  {43}},\ \bibinfo {pages} {5950} (\bibinfo {year} {1991})}\BibitemShut
  {NoStop}%
\bibitem [{\citenamefont {Sylju\aa{}sen}\ and\ \citenamefont
  {Sandvik}(2002)}]{Syljuasen2002}%
  \BibitemOpen
  \bibfield  {author} {\bibinfo {author} {\bibfnamefont {O.~F.}\ \bibnamefont
  {Sylju\aa{}sen}}\ and\ \bibinfo {author} {\bibfnamefont {A.~W.}\ \bibnamefont
  {Sandvik}},\ }\href {\doibase 10.1103/PhysRevE.66.046701} {\bibfield
  {journal} {\bibinfo  {journal} {Phys. Rev. E}\ }\textbf {\bibinfo {volume}
  {66}},\ \bibinfo {pages} {046701} (\bibinfo {year} {2002})}\BibitemShut
  {NoStop}%
\bibitem [{\citenamefont {Sylju\aa{}sen}(2003)}]{Syljuasen2003}%
  \BibitemOpen
  \bibfield  {author} {\bibinfo {author} {\bibfnamefont {O.~F.}\ \bibnamefont
  {Sylju\aa{}sen}},\ }\href {\doibase 10.1103/PhysRevE.67.046701} {\bibfield
  {journal} {\bibinfo  {journal} {Phys. Rev. E}\ }\textbf {\bibinfo {volume}
  {67}},\ \bibinfo {pages} {046701} (\bibinfo {year} {2003})}\BibitemShut
  {NoStop}%
\bibitem [{\citenamefont {Sandvik}(1997)}]{Sandvik1997}%
  \BibitemOpen
  \bibfield  {author} {\bibinfo {author} {\bibfnamefont {A.~W.}\ \bibnamefont
  {Sandvik}},\ }\href {\doibase 10.1103/PhysRevB.56.11678} {\bibfield
  {journal} {\bibinfo  {journal} {Phys. Rev. B}\ }\textbf {\bibinfo {volume}
  {56}},\ \bibinfo {pages} {11678} (\bibinfo {year} {1997})}\BibitemShut
  {NoStop}%
\bibitem [{\citenamefont {Sandvik}(1998)}]{Sandvik1998}%
  \BibitemOpen
  \bibfield  {author} {\bibinfo {author} {\bibfnamefont {A.~W.}\ \bibnamefont
  {Sandvik}},\ }\href {\doibase 10.1103/PhysRevLett.80.5196} {\bibfield
  {journal} {\bibinfo  {journal} {Phys. Rev. Lett.}\ }\textbf {\bibinfo
  {volume} {80}},\ \bibinfo {pages} {5196} (\bibinfo {year}
  {1998})}\BibitemShut {NoStop}%
\bibitem [{Wie()}]{Wierschem2014}%
  \BibitemOpen
  \href@noop {} {}\bibinfo {note} {See Supplemental Material for details of our
  projective QMC implementation}\BibitemShut {NoStop}%
\bibitem [{\citenamefont {Bera}\ \emph {et~al.}(2013)\citenamefont {Bera},
  \citenamefont {Lake}, \citenamefont {Islam}, \citenamefont {Klemke},
  \citenamefont {Faulhaber},\ and\ \citenamefont {Law}}]{Bera2013}%
  \BibitemOpen
  \bibfield  {author} {\bibinfo {author} {\bibfnamefont {A.~K.}\ \bibnamefont
  {Bera}}, \bibinfo {author} {\bibfnamefont {B.}~\bibnamefont {Lake}}, \bibinfo
  {author} {\bibfnamefont {A.~T. M.~N.}\ \bibnamefont {Islam}}, \bibinfo
  {author} {\bibfnamefont {B.}~\bibnamefont {Klemke}}, \bibinfo {author}
  {\bibfnamefont {E.}~\bibnamefont {Faulhaber}}, \ and\ \bibinfo {author}
  {\bibfnamefont {J.~M.}\ \bibnamefont {Law}},\ }\href {\doibase
  10.1103/PhysRevB.87.224423} {\bibfield  {journal} {\bibinfo  {journal} {Phys.
  Rev. B}\ }\textbf {\bibinfo {volume} {87}},\ \bibinfo {pages} {224423}
  (\bibinfo {year} {2013})}\BibitemShut {NoStop}%
\bibitem [{\citenamefont {Campostrini}\ \emph {et~al.}(2002)\citenamefont
  {Campostrini}, \citenamefont {Hasenbusch}, \citenamefont {Pelissetto},
  \citenamefont {Rossi},\ and\ \citenamefont {Vicari}}]{Campostrini2002}%
  \BibitemOpen
  \bibfield  {author} {\bibinfo {author} {\bibfnamefont {M.}~\bibnamefont
  {Campostrini}}, \bibinfo {author} {\bibfnamefont {M.}~\bibnamefont
  {Hasenbusch}}, \bibinfo {author} {\bibfnamefont {A.}~\bibnamefont
  {Pelissetto}}, \bibinfo {author} {\bibfnamefont {P.}~\bibnamefont {Rossi}}, \
  and\ \bibinfo {author} {\bibfnamefont {E.}~\bibnamefont {Vicari}},\ }\href
  {\doibase 10.1103/PhysRevB.65.144520} {\bibfield  {journal} {\bibinfo
  {journal} {Phys. Rev. B}\ }\textbf {\bibinfo {volume} {65}},\ \bibinfo
  {pages} {144520} (\bibinfo {year} {2002})}\BibitemShut {NoStop}%
\bibitem [{\citenamefont {Sandvik}\ and\ \citenamefont
  {Sudb\o}(1996)}]{Sandvik1996}%
  \BibitemOpen
  \bibfield  {author} {\bibinfo {author} {\bibfnamefont {A.~W.}\ \bibnamefont
  {Sandvik}}\ and\ \bibinfo {author} {\bibfnamefont {A.}~\bibnamefont
  {Sudb\o}},\ }\href {\doibase doi:10.1209/epl/i1996-00249-7} {\bibfield
  {journal} {\bibinfo  {journal} {Europhys. Lett.}\ }\textbf {\bibinfo {volume}
  {36}},\ \bibinfo {pages} {443} (\bibinfo {year} {1996})}\BibitemShut
  {NoStop}%
\bibitem [{\citenamefont {Wang}\ and\ \citenamefont
  {Sandvik}(2006)}]{Wang2006}%
  \BibitemOpen
  \bibfield  {author} {\bibinfo {author} {\bibfnamefont {L.}~\bibnamefont
  {Wang}}\ and\ \bibinfo {author} {\bibfnamefont {A.~W.}\ \bibnamefont
  {Sandvik}},\ }\href {\doibase 10.1103/PhysRevLett.97.117204} {\bibfield
  {journal} {\bibinfo  {journal} {Phys. Rev. Lett.}\ }\textbf {\bibinfo
  {volume} {97}},\ \bibinfo {pages} {117204} (\bibinfo {year}
  {2006})}\BibitemShut {NoStop}%
\bibitem [{\citenamefont {Zheludev}\ \emph {et~al.}(2001)\citenamefont
  {Zheludev}, \citenamefont {Chen}, \citenamefont {Broholm}, \citenamefont
  {Honda},\ and\ \citenamefont {Katsumata}}]{Zheludev2001}%
  \BibitemOpen
  \bibfield  {author} {\bibinfo {author} {\bibfnamefont {A.}~\bibnamefont
  {Zheludev}}, \bibinfo {author} {\bibfnamefont {Y.}~\bibnamefont {Chen}},
  \bibinfo {author} {\bibfnamefont {C.~L.}\ \bibnamefont {Broholm}}, \bibinfo
  {author} {\bibfnamefont {Z.}~\bibnamefont {Honda}}, \ and\ \bibinfo {author}
  {\bibfnamefont {K.}~\bibnamefont {Katsumata}},\ }\href {\doibase
  10.1103/PhysRevB.63.104410} {\bibfield  {journal} {\bibinfo  {journal} {Phys.
  Rev. B}\ }\textbf {\bibinfo {volume} {63}},\ \bibinfo {pages} {104410}
  (\bibinfo {year} {2001})}\BibitemShut {NoStop}%
\bibitem [{\citenamefont {Zheludev}\ \emph {et~al.}(2004)\citenamefont
  {Zheludev}, \citenamefont {Shapiro}, \citenamefont {Honda}, \citenamefont
  {Katsumata}, \citenamefont {Grenier}, \citenamefont {Ressouche},
  \citenamefont {Regnault}, \citenamefont {Chen}, \citenamefont {Vorderwisch},
  \citenamefont {Mikeska},\ and\ \citenamefont {Kolezhuk}}]{Zheludev2004}%
  \BibitemOpen
  \bibfield  {author} {\bibinfo {author} {\bibfnamefont {A.}~\bibnamefont
  {Zheludev}}, \bibinfo {author} {\bibfnamefont {S.~M.}\ \bibnamefont
  {Shapiro}}, \bibinfo {author} {\bibfnamefont {Z.}~\bibnamefont {Honda}},
  \bibinfo {author} {\bibfnamefont {K.}~\bibnamefont {Katsumata}}, \bibinfo
  {author} {\bibfnamefont {B.}~\bibnamefont {Grenier}}, \bibinfo {author}
  {\bibfnamefont {E.}~\bibnamefont {Ressouche}}, \bibinfo {author}
  {\bibfnamefont {L.-P.}\ \bibnamefont {Regnault}}, \bibinfo {author}
  {\bibfnamefont {Y.}~\bibnamefont {Chen}}, \bibinfo {author} {\bibfnamefont
  {P.}~\bibnamefont {Vorderwisch}}, \bibinfo {author} {\bibfnamefont {H.-J.}\
  \bibnamefont {Mikeska}}, \ and\ \bibinfo {author} {\bibfnamefont {A.~K.}\
  \bibnamefont {Kolezhuk}},\ }\href {\doibase 10.1103/PhysRevB.69.054414}
  {\bibfield  {journal} {\bibinfo  {journal} {Phys. Rev. B}\ }\textbf {\bibinfo
  {volume} {69}},\ \bibinfo {pages} {054414} (\bibinfo {year}
  {2004})}\BibitemShut {NoStop}%
\bibitem [{\citenamefont {den Nijs}\ and\ \citenamefont
  {Rommelse}(1989)}]{denNijs1989}%
  \BibitemOpen
  \bibfield  {author} {\bibinfo {author} {\bibfnamefont {M.}~\bibnamefont {den
  Nijs}}\ and\ \bibinfo {author} {\bibfnamefont {K.}~\bibnamefont {Rommelse}},\
  }\href {\doibase 10.1103/PhysRevB.40.4709} {\bibfield  {journal} {\bibinfo
  {journal} {Phys. Rev. B}\ }\textbf {\bibinfo {volume} {40}},\ \bibinfo
  {pages} {4709} (\bibinfo {year} {1989})}\BibitemShut {NoStop}%
\bibitem [{\citenamefont {Anfuso}\ and\ \citenamefont
  {Rosch}(2007{\natexlab{a}})}]{Anfuso2007a}%
  \BibitemOpen
  \bibfield  {author} {\bibinfo {author} {\bibfnamefont {F.}~\bibnamefont
  {Anfuso}}\ and\ \bibinfo {author} {\bibfnamefont {A.}~\bibnamefont {Rosch}},\
  }\href {\doibase 10.1103/PhysRevB.75.144420} {\bibfield  {journal} {\bibinfo
  {journal} {Phys. Rev. B}\ }\textbf {\bibinfo {volume} {75}},\ \bibinfo
  {pages} {144420} (\bibinfo {year} {2007}{\natexlab{a}})}\BibitemShut
  {NoStop}%
\bibitem [{\citenamefont {Anfuso}\ and\ \citenamefont
  {Rosch}(2007{\natexlab{b}})}]{Anfuso2007b}%
  \BibitemOpen
  \bibfield  {author} {\bibinfo {author} {\bibfnamefont {F.}~\bibnamefont
  {Anfuso}}\ and\ \bibinfo {author} {\bibfnamefont {A.}~\bibnamefont {Rosch}},\
  }\href {\doibase 10.1103/PhysRevB.76.085124} {\bibfield  {journal} {\bibinfo
  {journal} {Phys. Rev. B}\ }\textbf {\bibinfo {volume} {76}},\ \bibinfo
  {pages} {085124} (\bibinfo {year} {2007}{\natexlab{b}})}\BibitemShut
  {NoStop}%
\bibitem [{\citenamefont {You}\ \emph {et~al.}(2013)\citenamefont {You},
  \citenamefont {Bi}, \citenamefont {Rasmussen}, \citenamefont {Slagle},\ and\
  \citenamefont {Xu}}]{You2013}%
  \BibitemOpen
  \bibfield  {author} {\bibinfo {author} {\bibfnamefont {Y.-Z.}\ \bibnamefont
  {You}}, \bibinfo {author} {\bibfnamefont {Z.}~\bibnamefont {Bi}}, \bibinfo
  {author} {\bibfnamefont {A.}~\bibnamefont {Rasmussen}}, \bibinfo {author}
  {\bibfnamefont {K.}~\bibnamefont {Slagle}}, \ and\ \bibinfo {author}
  {\bibfnamefont {C.}~\bibnamefont {Xu}},\ }\href@noop {} {} (\bibinfo {year}
  {2013}),\ \bibinfo {note} {arXiv:1312.0626}\BibitemShut {NoStop}%
\bibitem [{\citenamefont {Todo}\ \emph {et~al.}(2001)\citenamefont {Todo},
  \citenamefont {Matsumoto}, \citenamefont {Yasuda},\ and\ \citenamefont
  {Takayama}}]{Todo2001}%
  \BibitemOpen
  \bibfield  {author} {\bibinfo {author} {\bibfnamefont {S.}~\bibnamefont
  {Todo}}, \bibinfo {author} {\bibfnamefont {M.}~\bibnamefont {Matsumoto}},
  \bibinfo {author} {\bibfnamefont {C.}~\bibnamefont {Yasuda}}, \ and\ \bibinfo
  {author} {\bibfnamefont {H.}~\bibnamefont {Takayama}},\ }\href {\doibase
  10.1103/PhysRevB.64.224412} {\bibfield  {journal} {\bibinfo  {journal} {Phys.
  Rev. B}\ }\textbf {\bibinfo {volume} {64}},\ \bibinfo {pages} {224412}
  (\bibinfo {year} {2001})}\BibitemShut {NoStop}%
\bibitem [{\citenamefont {Charrier}\ \emph {et~al.}(2010)\citenamefont
  {Charrier}, \citenamefont {Capponi}, \citenamefont {Oshikawa},\ and\
  \citenamefont {Pujol}}]{Charrier2010}%
  \BibitemOpen
  \bibfield  {author} {\bibinfo {author} {\bibfnamefont {D.}~\bibnamefont
  {Charrier}}, \bibinfo {author} {\bibfnamefont {S.}~\bibnamefont {Capponi}},
  \bibinfo {author} {\bibfnamefont {M.}~\bibnamefont {Oshikawa}}, \ and\
  \bibinfo {author} {\bibfnamefont {P.}~\bibnamefont {Pujol}},\ }\href
  {\doibase 10.1103/PhysRevB.82.075108} {\bibfield  {journal} {\bibinfo
  {journal} {Phys. Rev. B}\ }\textbf {\bibinfo {volume} {82}},\ \bibinfo
  {pages} {075108} (\bibinfo {year} {2010})}\BibitemShut {NoStop}%
\end{thebibliography}%

\section{Supplementary Material: Projective Quantum Monte Carlo}

To investigate the strange correlator of You {\it et al.}~\cite{You2013}, we use a projective variant of the stochastic series expansion (SSE) quantum Monte Carlo method~\cite{Sandvik1991}. The main idea is as follows: instead of expanding the density matrix as a Taylor series of Hamiltonian operations, we project out the ground state by repeated Hamiltonian operation upon a trial wave function. While the presence of a trial wave function explicitly removes the usual periodicity in the imaginary-timelike dimension of the operator string, it can be thought of as a set of vertices of infinite weight. Thus, we can still utilize the directed loop equations of Sylju\r{a}sen and Sandvik~\cite{Syljuasen2002}, minimize bounce probabilities in the loop algorithm, and obtain efficient global updates.

Now we describe our projective quantum Monte Carlo scheme in detail. First, let us examine the effect of $m$ repeated Hamiltonian operations on a trial wave function,
\begin{equation}
\left({\cal H}-C\right)^{m}|\psi\rangle=\left({\cal H}-C\right)^{m}\sum_{\alpha}c_{\alpha}|\alpha\rangle.
\end{equation}
Here, we have expanded $|\psi\rangle$ in the basis of energy eigenstates $|\alpha\rangle$ with coefficients given by $c_{\alpha}=\langle\alpha|\psi\rangle$. The constant $C$ is chosen to make $\left({\cal H}-C\right)$ negative definite. Thus, as long as $c_{0}\neq0$, the projection of $|\psi\rangle$ will be dominated by the ground state terms,
\begin{equation}
\left({\cal H}-C\right)^{m}|\psi\rangle=\sum_{\alpha}c_{\alpha}\left(E_{\alpha}-C\right)^{m}|\alpha\rangle.
\end{equation}
This can be made more explicit by rewriting the expression as
\begin{equation}
\left(\frac{{\cal H}-C}{E_{0}-C}\right)^{m}|\psi\rangle=\sum_{\alpha}c_{\alpha}\left(\frac{E_{\alpha}-C}{E_{0}-C}\right)^{m}|\alpha\rangle.
\end{equation}
Thus, the ground state is approached as $m\rightarrow\infty$. Having a valid ground state projector, we can evaluate ground state observables as
\begin{equation}
\langle{\cal O}\rangle=\frac{\langle\psi|\left({\cal H}-C\right)^{m}{\cal O}\left({\cal H}-C\right)^{m}|\psi\rangle}{\langle\psi|\left({\cal H}-C\right)^{2m}|\psi\rangle}.
\end{equation}
Within the same formulation, we can also easily compute overlap of the ground state wave function with an arbitrary wave function $|\Omega\rangle$, as required for calculating the so-called strange correlator~\cite{You2013}
\begin{equation}
C_{SC}(i,j)=\frac{\langle\Omega|S_i^+S_j^-\left({\cal H}-C\right)^{2m}|\psi\rangle}{\langle\Omega|\left({\cal H}-C\right)^{2m}|\psi\rangle}.
\end{equation}

In our calculations, we choose
\begin{equation}
|\Omega\rangle=\prod_i|0\rangle_i\otimes
\end{equation}
and hence $|\Omega\rangle$ is a trivial direct product state, as required for calculating the strange correlator~\cite{You2013}. Without any loss of generality, we can set $|\psi\rangle=|\Omega\rangle$ for our trial wave function. By comparing the staggered magnetization and strange correlator with exact values for finite chains, we have verified the accuracy of our projective QMC method, as well as its ability to obtain ground state results. By analogy to traditional SSE, the length of the operator string $m$ required to reach the ground state limit is expected to scale as $L^{d+z}$, with $d$ the dimensionality and $z$ the dynamic critical exponent. We have found $m=L^3/4$ to be sufficient to reach the ground state limit for the results presented in Fig.~5 of the main text.

\end{document}